\documentclass[runningheads]{llncs}

\usepackage[T1]{fontenc}
\usepackage{amsmath}
\usepackage{graphicx}
\usepackage{xcolor} 
\usepackage{booktabs}
\usepackage{array}
\usepackage{multirow}
\usepackage{tabularx}  
\usepackage{xurl}
\usepackage{comment}

\begin{document}
\begin{sloppypar}
    
\title{Building the Truman Show: A TrustZone-Based Framework for Lightweight Out-of-band Kernel Security Monitoring}

\titlerunning{A Lightweight Out-of-band OS Monitor for Kernel Security}

\author{Zhenling Duan\inst{1}\orcidID{0009-0004-7097-2494} \and
Pan Dong\inst{1}\thanks{Corresponding author:pandong@nudt.edu.cn} \orcidID{0000-0002-2890-260X} \and
Renshuang Jiang\inst{1}\orcidID{0009-0002-4411-8150} \and
Xiaoxiang Fang\inst{1}\orcidID{0009-0000-5409-8670} \and
Bao Li\inst{1}\orcidID{0000-0002-4504-6841} }

\authorrunning{Z. Duan et al.}

\institute{College of Computer Science and Technology, National University of Defense Technology, Changsha, China
\\
\email{
\{dzl,pandong,rshuang,baoli\}@nudt.edu.cn
}
}

\maketitle             

\begin{abstract}
The increasing number of vulnerabilities in operating systems, together with sophisticated kernel-level threats (e.g., rootkits), has weakened the effectiveness of traditional in-kernel protection mechanisms. Since these defenses operate at the same privilege level as the kernel, they share the same attack surface and can be bypassed once the kernel is compromised. Isolation-based security approaches provide stronger protection by separating security logic from the kernel, but strict isolation often introduces semantic gaps that limit system visibility and hinder timely threat detection.
In this paper, we present LOOM, a lightweight out-of-band operating system monitoring architecture built on ARM TrustZone. By leveraging TrustZone’s hardware-enforced isolation, LOOM establishes a tamper-resistant monitoring environment independent of the kernel. To bridge the semantic gap, we design a lightweight semantic reconstruction mechanism in the Secure World. It selectively captures the states and behavioral patterns of critical kernel objects, such as process control blocks and kernel modules. Additionally, LOOM introduces a dual-stage hazard prevention mechanism that combines atomic memory protection with an interrupt-driven adaptive agent to detect and mitigate kernel rootkit activities. An address translation cache is further incorporated to optimize repeated address access and reduce monitoring overhead. Overall, we develop a multi-layered collaborative architecture with platform, functional, and auxiliary layers for secure and efficient kernel monitoring. A prototype of LOOM has been implemented on the Phytium D2000 platform. Experimental results indicate that LOOM incurs negligible overhead while maintaining a strong monitoring capability. Furthermore, a security capability analysis based on CVE cases demonstrates that LOOM can detect and mitigate various kernel attacks. 
\keywords{ARM TrustZone \and Out-of-Band monitoring \and Kernel security  \and Threats detection \and Isolation mechanisms.}

\end{abstract}

\section{Introduction}
The increasing prevalence of vulnerabilities within operating system (OS) kernels, coupled with kernel-level threats such as rootkits~\cite{ref1}, is diminishing the efficacy of traditional protection methods.

Conventional approaches are usually implemented within the system kernel, where protective measures and attack targets operate at the same privilege level. This overlap in attack surfaces creates a shared-fate dependency: once the kernel is compromised, the protection mechanisms also fail. Given the inherent complexity and fragility of kernel structures, some in-kernel mechanisms are susceptible to vulnerabilities such as privilege escalation and control-flow hijacking, which further challenge the reliability.

To overcome these limitations, an isolation mechanism that provides OS-independent security enforcement and remains resilient to kernel interference offers a promising solution.  Rather than strengthening kernel-resident defenses, we advocate for an out-of-band monitoring design that operates independently, ensuring reliable security even if the kernel is compromised.

Isolation can be achieved through various infrastructures, including pure hardware isolation (e.g., Trusted Platform Module (TPM)~\cite{ref2} and Hardware Security Module (HSM)~\cite{ref3}), software virtualization isolation (e.g., microkernels and containers), or hardware-software collaborative isolation (e.g., Trusted Execution Environment (TEE)~\cite{ref4}). However, these technologies trade off in terms of security, perceptual capabilities, and performance overhead. For instance, while hardware isolation provides superior security, it can limit system flexibility. On the other hand, software isolation is generally easier to deploy but may introduce additional vulnerabilities.

Furthermore, increased isolation creates larger semantic gaps. Limited interaction between security components and the system kernel reduces visibility and understanding of the system state, while raising observation and enforcement costs. Thus, effective out-of-band security monitoring within an isolation framework faces some critical, unresolved challenges:
\begin{itemize}
\item[1)]Hardware-enforced isolation significantly restricts the secure world’s understanding of kernel semantics, making it inherently difficult to achieve accurate system comprehension without compromising isolation guarantees.
\item[2)]Semantic reconstruction faces a dilemma between completeness and complexity.  Comprehensive kernel mirroring is unfeasible and expands the attack surface, while inadequate semantics diminish the effectiveness of monitoring.
\item[3)]Out-of-band monitoring presents challenges to timely threat detection and response, as it relies on indirect observation and mediated control instead of direct involvement with the kernel.
\item[4)]Balancing monitoring overhead with security is challenging, as aggressive optimization can compromise guarantees or introduce new trust assumptions.
\end{itemize}

This progression marks a shift from "kernel-reliant protection" to "isolation-based security," highlighting the tension between isolation and information accessibility. To address this, we propose a lightweight out-of-band OS monitor (LOOM) based on the ARM TrustZone.By leveraging TrustZone’s hardware-enforced isolation, which provides a separate execution domain from the Normal World, our approach establishes a tamper-resistant security foundation.     Additionally, it enables high-speed, low-intrusion memory operations in the Normal World through its built-in one-way memory access mechanism and high-privilege register interface.
This design facilitates the detection and prevention of kernel rootkits that attempt to manipulate kernel objects or execution paths while evading detection by in-kernel monitors. Consequently, this framework is an ideal platform for effective out-of-band monitoring.

Building upon this foundation, a lightweight semantic reconstruction module is integrated within the Secure World to dynamically monitor and capture the state and behavior patterns of key kernel objects and events. The monitored Objects include Process Control Blocks (task\_struct), File Descriptor Tables (files\_operations), kernel module lists, and so on. Additionally, a dual-stage hazard prevention is introduced. First, atomic operations mark certain pages as read-only or non-executable to protect the page table from tampering. Second, an adaptive agent uses interrupts to swiftly detect and halt kernel rootkits when it identifies malicious activity. Moreover, we introduce an address translation cache in LOOM to optimize overhead.

We adopt a multi-layered collaborative architecture consisting of (i) a platform layer that provides TrustZone-based isolation and the channel of access, (ii) a functional layer responsible for semantics-aware monitoring and behavioral control, and (iii) an auxiliary layer that optimizes the system performance overhead. Together, these layers form a secure, controllable, and efficient LOOM for kernel security. The primary contributions of this work are as follows:

\begin{itemize}
\item[$\bullet$] We introduce the LOOM framework, which facilitates effective monitoring of kernel security and adds the capability to prevent emerging threats, thereby filling the gap in secure kernel monitoring based on the ARM TrustZone.
\item[$\bullet$] We have developed a prototype on the Phytium D2000, and our experimental results show that it brings negligible overhead.
\item[$\bullet$] The security capabilities of LOOM are analyzed through representative CVE case studies, demonstrating the types of vulnerabilities that can be effectively detected.
\end{itemize}

\section{Related Work}

\subsection{Isolation Mechanism}

Previous research on isolation mechanisms has mainly focused on separating trust-critical functions from the monolithic kernel, often using hardware-based solutions such as TPMs and HSMs~\cite{ref2,ref3}. These devices enhance system trust by supporting measured boot processes and secure key storage. However, hardware solutions have limitations in terms of programmability and applicability, making them unsuitable for continuous OS-level monitoring.  In contrast, software-based isolation approaches, including microkernels (e.g., seL4~\cite{ref9}), aim to reduce the trusted computing base by moving services to user space. In addition, hardware-assisted virtualization (e.g., KVM~\cite{ref10}) isolates entire OS instances. These approaches introduce additional system complexity and expand the trusted computing base to include the microkernel or hypervisor, which may itself become an attack target. To balance security and practicality, Collaborative hardware–software TEEs like Intel SGX~\cite{ref11}, ARM TrustZone~\cite{ref5}, and AMD SEV~\cite{ref12}, provide isolated execution domains for protecting sensitive code and data through mechanisms such as secure memory regions and sealed storage.

\begin{figure}
\centering
\includegraphics[width=0.75\textwidth]{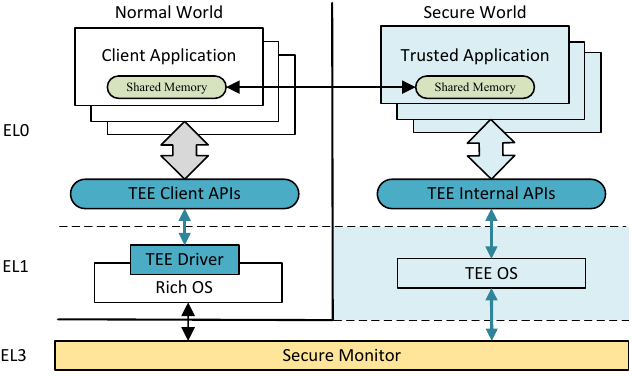}
\caption{ARMv8-A TrustZone software architecture} \label{fig1}
\end{figure}

Especially, ARM TrustZone~\cite{ref5} is a hardware-based security extension designed for ARM processors, primarily aimed at safeguarding sensitive code and data. The software architecture of ARMv8-A TrustZone is illustrated in Fig.~\ref{fig1}. At the hardware level, ARMv8-A defines four exception levels (EL0–EL3) and two security states (Secure and Non-secure). Each physical core is virtualized into two logical cores, distinguished by a Non-Secure (NS) bit on the AXI bus, which facilitates isolation across the processor core, memory, and peripherals. In the software layer, the architecture includes the Rich Execution Environment (REE) and the TEE~\cite{ref4}. The Secure Monitor, operating at EL3, is tasked with managing world switching and state transitions. Various studies have leveraged ARM TrustZone to enhance both security and performance. For example, TZ-IMA~\cite{ref6} focuses on integrity measurement, whereas TZDKS~\cite{ref7} introduces a dual-criticality system that balances performance with security. Furthermore, Li et al.~\cite{ref8} investigated performance prediction and mandatory access control as additional means to strengthen security.

\subsection{out-of-band monitoring}

Out-of-band monitoring decouples system observation from the OS kernel, reducing the attack surface and improving resilience against kernel compromise. Pure hardware-based monitors provide strong isolation but limited visibility. For example, Hypernel’s Memory Bus Monitor (MBM)~\cite{ref13} detects anomalies by observing memory bus traffic, yet it lacks access to CPU-internal states and dynamically allocated kernel objects, limiting its ability to capture fine-grained kernel behaviors.
Virtual Machine Introspection (VMI)~\cite{ref14} improves visibility by relocating monitoring to a privileged domain outside guest virtual machines, enabling access to low-level system state. Systems such as NFM~\cite{ref15} and VMI IDS~\cite{ref16} detect rootkits by reconstructing high-level kernel semantics from low-level system states(such as guest memory and CPU state). However, VMI relies on software-based isolation. Its security depends on the correctness of the hypervisor and may inherit vulnerabilities from the virtualization layer.
Hybrid approaches attempt to combine hardware-assisted isolation with software to balance these limitations. 
For instance, KIMS~\cite{ref17}leverages ARM TrustZone to enhance kernel integrity measurement, but it still suffers from limited semantic understanding and inefficient interaction with the monitored kernel.

\section{Design}
In this section, we first describe the four design principles and the overall architecture of LOOM (Section 3.1), then present its threat model (Section 3.2).

\subsection{Framework}

In network management, in-band management utilizes standard protocols like Telnet or SSH, where management traffic shares the same path as production traffic. This may lead to complications during network disruptions. In such scenarios, out-of-band management is proposed to provide a separate channel for device access, ensuring functionality even when the primary data path is down.

Similarly, the idea also applies to OS monitoring. In-band monitoring mechanisms operate within the same trust domain as the monitored kernel, typically running in the Normal World of kernel mode. And they share the same virtual address space and privilege level as the OS, without an independent root of trust. In contrast, an out-of-band monitoring model places the monitoring logic in an isolated trust domain, separated from the potentially compromised operating system by a hardware-enforced boundary. In the case of LOOM, this trusted domain is the Secure World provided by Arm TrustZone. The monitoring logic executes in the Secure World, while the kernel operates in the Normal World. The architectural isolation between the two worlds is enforced by hardware, ensuring that the monitoring component is not directly controlled by the normal-world kernel. This separation establishes a stronger trust foundation and enhances resilience against kernel-level compromises.

To better tackle the challenges outlined in the introduction, LOOM should be designed in accordance with the following four design principles by integrating the hardware isolation capabilities of ARM TrustZone.

\begin{itemize}
{\item[$\bullet$]\textbf{Isolation–awareness trade-off under hardware constraints:}Hardware isolation limits the ability of the secure world to access/understand resources, particularly in obtaining kernel information. Effective kernel monitoring requires the secure world to understand kernel-level states. Thus, it is essential to allow the secure world to access necessary kernel information while preserving the hardware security boundary.
\item[$\bullet$] \textbf{Targeted and efficient semantic reconstruction:}Reconstructs a minimal subset of kernel semantics for monitoring and detection, to reduce system complexity and the attack surface.
\item[$\bullet$]\textbf{Timely threat detection and response:}Upon detection of anomalous kernel behavior, the monitoring mechanism triggers an immediate, targeted response to mitigate potential hazards.
\item[$\bullet$]\textbf{Optimized overhead without compromising security:}Performance optimizations must conform to strict requirements regarding non-interference and correctness; no additional trust assumptions are introduced.}
\end{itemize}

\begin{figure}
\centering
\includegraphics[width=0.83\textwidth]{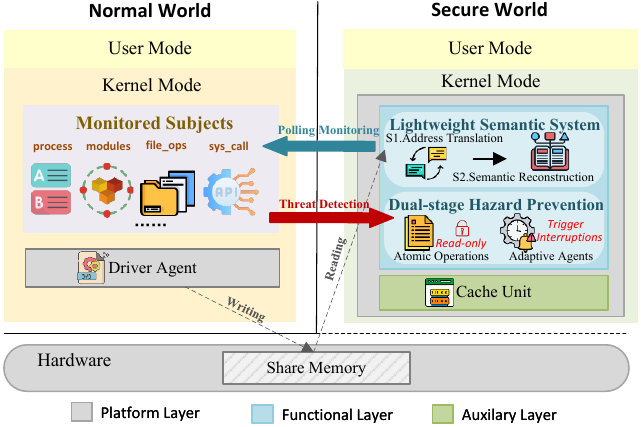}
\caption{Architecture of LOOM} \label{fig2}
\end{figure}

As shown in Fig.~\ref{fig2}, LOOM employs a layered architecture comprising a platform layer, a functional layer, and an auxiliary layer. This architecture enables effective kernel security monitoring and threat response. The monitoring logic is located in the functional layer within the Secure World. This setup effectively separates monitoring from the Normal World kernel, which could be vulnerable to compromise. The monitored subjects are kernel entities in the Normal World that are susceptible to rootkit attacks and carry critical kernel information, such as processes, the system call table, and modules. For instance, if malware disguises itself as a kernel loadable module(KLM), the monitoring component can spot it by analyzing the module list of the kernel.

\textbf{Platform Layer.}
It relies on the hardware-enforced isolation of ARM TrustZone as the foundation for out-of-band monitoring. A driver agent statically compiled into the Normal World kernel exports selected security-critical kernel objects whose locations are defined in the \texttt{System. map}(a symbol table file in the Linux kernel). The agent performs read-only access to these kernel data structures and forwards their states to the Secure World through a static shared memory region.
During secure boot, this memory region is initialized as a unidirectional channel and mapped read-only in the Secure World, where it serves solely as a passive data buffer without executable code or control logic. After initialization, the Secure World retrieves the states of security-critical kernel objects from the shared memory to establish a trusted baseline for monitoring.
Afterwards, the Secure World considers the forwarded data as untrusted input, while retaining exclusive control over the monitoring logic and analysis procedures. Even if an attacker tampers with the buffer contents at runtime, such manipulation cannot interfere with the monitoring logic. Instead, inconsistencies introduced by forged data are exposed during LOOM’s semantic analysis. Once an anomaly is detected, the hazard prevention mechanism is immediately triggered to mitigate the potential threat.

\textbf{Functional Layer.}
It embodies LOOM’s core monitoring capabilities. A lightweight semantic reconstruction engine maps raw physical memory to a small set of security-critical kernel objects, including task lists, module lists, file system objects, and system call contexts.  Rather than reconstructing full kernel semantics, LOOM focuses only on essential objects, thereby reducing complexity and attack surface.  Reconstruction is performed entirely in the Secure World and without kernel cooperation.
Moreover, hazard prevention follows a dual-stage design: atomic operations enforce read-only or non-executable protections on critical page-table pages, while an adaptive interrupt-driven agent can rapidly detect and halt kernel rootkits.

\textbf{Auxiliary Layer.}
It reduces the cost of semantic reconstruction by caching translations between physical memory addresses and their Secure World virtual mappings. By avoiding repeated address translation for frequently accessed kernel objects, this layer significantly reduces monitoring latency and overhead.

The LOOM workflow proceeds as follows. A driver in the REE kernel extracts monitored object information from the \texttt{System.map} and writes the state to a static shared memory region. The LOOM retrieves this data and periodically reconstructs kernel object semantics to analyze security-critical attributes. Each monitoring task takes about 100 ms on average, including address resolution and semantic reconstruction. LOOM operates at a polling frequency of 10 Hz. When semantic inconsistencies or invalid properties are detected, hazard prevention is immediately triggered.

\subsection{Threat Model}
LOOM assumes the correctness and trustworthiness of the ARM TrustZone hardware, the secure boot chain, and all Secure World software components, collectively forming the trusted computing base (TCB). These trusted components enforce hardware-level isolation between the Secure World and the Normal World.
The adversary is assumed to have full control over the Normal World kernel. Specifically, the attacker can modify kernel code and data, manipulate kernel control structures (e.g., process lists and system call tables), dynamically load malicious modules, and conceal their presence by tampering with kernel data structures. However, the attacker is assumed to be incapable of violating TrustZone isolation, executing code in the Secure World, or modifying Secure World memory.
To minimize additional attack surface, the kernel driver and shared memory channel are designed to be minimal and passive. They perform only deterministic export of raw kernel state through a static hardware-protected shared memory region and do not enforce security policies or influence kernel control flow. The Secure World treats all exported data as untrusted input and performs cross-validation and invariant checking to detect malicious manipulation.

Instead of aiming for broad-spectrum detection of arbitrary kernel attacks, LOOM focuses on in-depth monitoring of security-critical kernel objects that are closely related to specific rootkit attack surfaces. For example, malicious KLM activities are detected by reconstructing and validating the semantics and integrity of module-related kernel data structures. The scope of the system excludes physical hardware attacks (e.g., cold boot attacks  ~\cite{ref18}, or bus snooping  ~\cite{ref19}), side-channel attacks such as PACMAN  ~\cite{ref20}, or Denial-of-Service (DoS) that aim to exhaust system resources.

\section{Implementation}
This section describes the implementation of LOOM and its key components. We first present the lightweight out-of-band semantic reconstruction mechanism used to recover kernel state (Section 4.1). Then we describe the dual-stage hazard prevention mechanism for rapid threat mitigation (Section 4.2). Finally, we introduce a caching mechanism to reduce monitoring overhead (Section 4.3).

\subsection{Lightweight Out-of-Band Semantic Reconstruction}

\begin{figure}
\centering
\includegraphics[width=\textwidth]{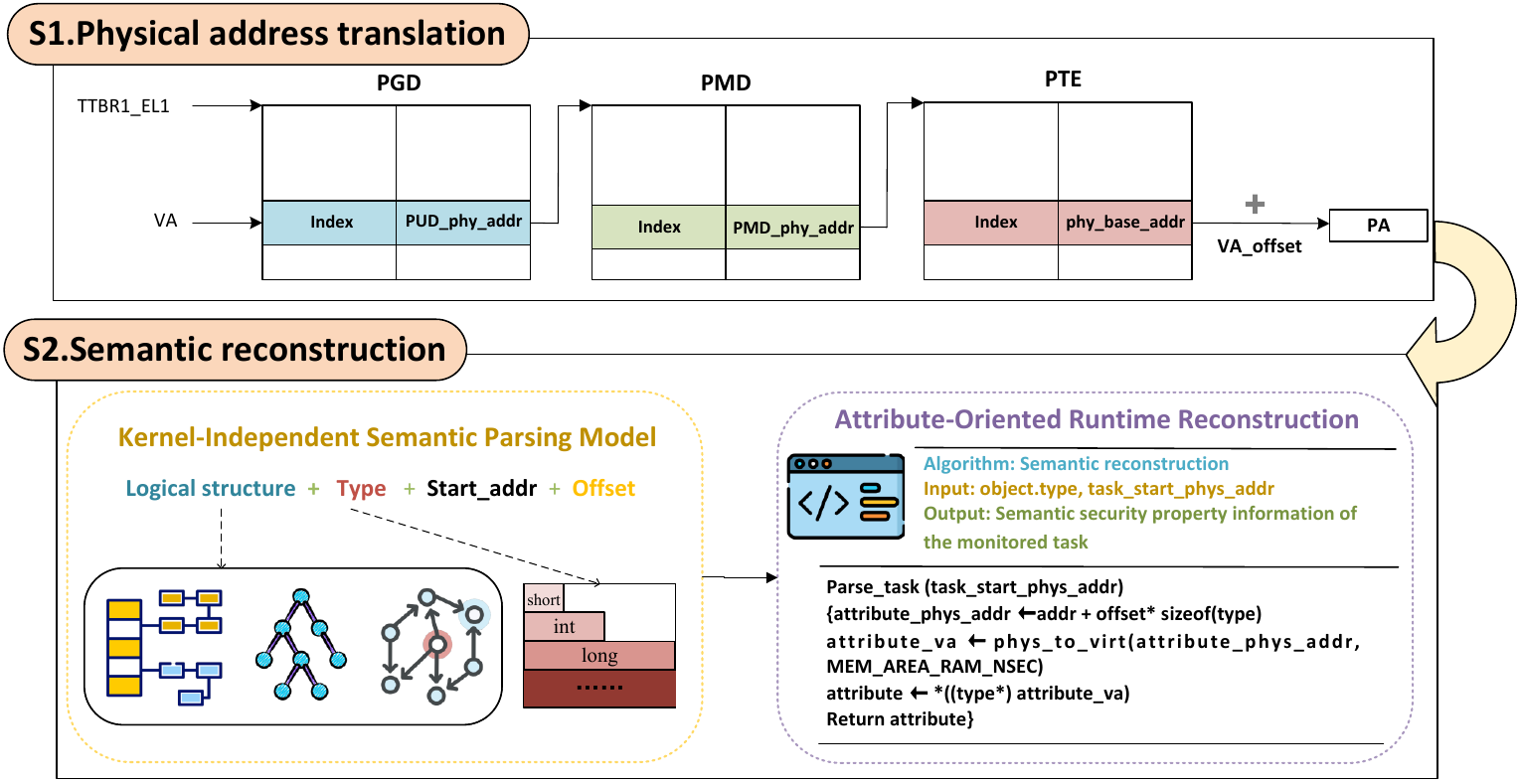}
\caption{Procedure of the semantic reconstruction} \label{fig3}
\end{figure}

Independent monitoring of kernel data objects is realized within the Secure World through a lightweight out-of-band semantic system. The system consists of two tightly coupled stages: S1) physical address resolution and S2) semantic reconstruction, as shown in Fig.~\ref{fig3}. In this design, physical addresses are obtained through a step-by-step traversal of the hierarchical page table structure. Rather than introducing new address-translation mechanisms, the contribution lies in how kernel semantics are reconstructed and reasoned about under strict isolation, which forms the core technical focus of this component.

\subsubsection{S1)Physical Address Translation.}

A physical address represents the actual location of data in memory. Although the REE and TEE share the same untrusted physical memory, they maintain independent page table mappings, allowing the OS to access the same memory through different virtual addresses. After boot, the REE kernel writes critical entries from \texttt{System.map} into a 12 MB shared memory region. As a result, LOOM running in the TEE initially observes the virtual addresses of the REE kernel. By reading the \texttt{TTBR1\_EL1} register, which stores the base address of the kernel page table in ARMv8, LOOM obtains the base address of the REE kernel page tables.

To access the page table mapping of the REE kernel within the TEE, the shared memory region must first be registered and mapped using the \texttt{register\_phy\_map()} function. LOOM then traverses the page tables level by level according to the page-table configuration, including the number of levels, page size, and virtual address width, to resolve virtual-to-physical address mappings. In our setup, as illustrated in Fig.~\ref{fig3} (S1), the system employs a three-level page table consisting of a Page Global Directory (PGD), a Page Middle Directory (PMD), and Page Table Entries (PTEs). This configuration uses 4 KB pages and supports a 39-bit virtual address space. Each level of the page table contains 512 entries ($2^{9}$). The PTE provides the base physical frame address, which, when combined with the page offset (\texttt{vaddr[12:0]}), yields the target physical address. Finally, the \texttt{phys\_to\_virt()} function enables safe access to this physical address from within the TEE.

By directly resolving the physical addresses of monitored objects within LOOM, the design effectively mitigates address-mapping deception that could result from malicious tampering with the REE kernel page tables. Furthermore, obtaining correct physical addresses is a prerequisite for subsequent semantic reconstruction, thereby enhancing the robustness of out-of-band monitoring.

\subsubsection{S2)Lightweight Semantic Reconstruction.}

Out-of-band kernel monitoring fundamentally suffers from the semantic gap between raw physical memory and high-level kernel semantics. Although the monitor can directly observe physical memory, the meaning of a memory word cannot be inferred without kernel-specific context: the same 64-bit value may represent a process identifier or a function pointer. This lack of semantic interpretation prevents direct reasoning about kernel behavior.
To bridge this gap, prior work~\cite{ref33,ref34} has relied on techniques such as in-kernel cooperation, symbol introspection, and full data structure reconstruction (e.g., VMI-based approaches~\cite{ref35}).
While effective, these solutions either require kernel instrumentation, depend on debug symbols, or incur substantial runtime overhead due to comprehensive structure traversal. 

In this work, we introduce a kernel-independent, attribute-oriented semantic reconstruction approach.  Rather than reconstructing complete kernel data structures, we externalize semantic interpretation rules into a compact knowledge base embedded within a resident monitor.  The monitor selectively reconstructs only those kernel attributes(e.g., process descriptors, system call table, and kernel modules) that are strictly required by the monitoring logic, avoiding full structural traversal and redundant semantic recovery. This design fundamentally shifts semantic restoration from a structure-centric paradigm to a minimal, attribute-driven parsing model.  Instead of iteratively walking complex kernel objects to recover holistic semantics, our approach extracts a bounded set of critical attributes and derives security-relevant interpretations directly from them. 

In contrast to prior work, by decoupling semantic knowledge from kernel implementation details, our approach does not depend on kernel instrumentation, debug symbols at runtime, or in-kernel agents, etc. 
During system initialization, LOOM reads only from the shared memory with essential kernel metadata (e.g., \texttt{System.map}). Rather than embedding fixed kernel addresses or hard-coded offsets into the monitor, this metadata is used to instantiate lightweight semantic descriptors that support runtime dynamic address resolution. Consequently, semantic reconstruction remains adaptive across different kernel instances while avoiding the rigidity of conventional fixed-offset approaches. As illustrated in Fig.~\ref{fig3}, Stage 2(S2) is built upon two key innovations. 

\textbf{Kernel-Independent Semantic Parsing Model.}Semantic knowledge is decoupled from the monitored kernel and formalized as a compact semantic descriptor.  Each monitored element is defined by a quadruple ⟨logical structure, type, base address, offset⟩, which serves as the minimal semantic description unit. By representing kernel objects as composable semantic descriptors rather than complete data structures, we avoid building pointer graphs or reconstructing global layouts. Only attributes strictly necessary for monitoring are modeled, significantly reducing the semantic footprint.
\begin{itemize}
\item[$\bullet$] Logical structure: captures the conceptual access pattern of the attribute, such as membership in a linked list (e.g., task lists), hierarchical containment (e.g., process-to-credential relationships), or indexed access in an array (e.g., the system call table). The tree and list structures shown in Figure 3 represent the logical data structure types used to describe kernel attributes. They abstract common kernel organization patterns, facilitating intuitive modeling of semantic access paths and dependency relationships during reconstruction.
\item[$\bullet$] Type: specifies the primitive data format and size of the attribute (e.g., integer, pointer, or function address), ensuring correct interpretation of raw memory values during reconstruction.
\item[$\bullet$] Base address: the absolute location of a known element in physical memory.
\item[$\bullet$] Offset: defines the relative position of the target attribute with respect to the base address, enabling precise localization of the desired field.
\end{itemize}

To improve portability across kernel versions, LOOM does not rely on statically hard-coded offsets. Instead, descriptor parameters are resolved during boot-time initialization and can be automatically generated offline using standard kernel build tools or lightweight extraction scripts. Moreover, the monitored kernel objects targeted by LOOM (e.g., syscall tables and task lists) exhibit relatively stable structural organization across kernel versions, allowing the semantic descriptors to remain compact while preserving practical compatibility. Although kernel layout variability cannot be eliminated, our design represents a trade-off between full generality and lightweight, low-overhead semantic recovery.

\textbf{Attribute-Oriented Runtime Reconstruction.} At runtime, the monitor reconstructs kernel semantics incrementally using physical memory observations. For each target attribute, its physical address is computed by combining the preloaded base address and offset. The address is then mapped into a Secure World–accessible virtual address via TrustZone memory management, and the semantic value is safely obtained through pointer dereferencing.
The reconstruction follows the pre-modeled logical access paths but touches only the minimal set of memory locations necessary to recover the target semantic property. This enables efficient cross-world semantic interpretation with low overhead, while maintaining strict isolation from the monitored kernel.

\subsection{Hazard Prevention}

\begin{figure}
\centering
\includegraphics[width=\textwidth]{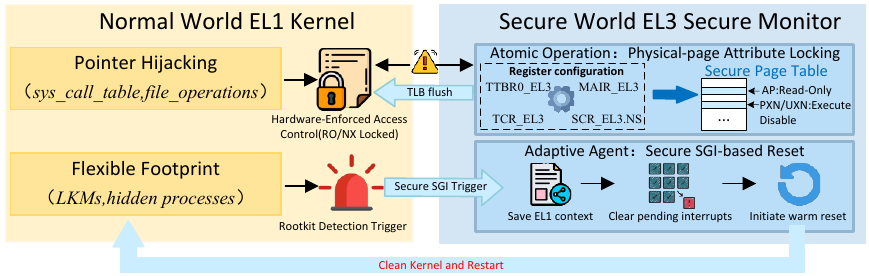}
\caption{Workflow of dual-stage hazard prevention} \label{fig5}
\end{figure}

To enhance the out-of-band response capability of LOOM and minimize the security impact of rootkits, monitoring information derived from semantic reconstruction is utilized to enable hazard prevention. All rootkits in this study are detectable through consistency checks of critical semantic attributes or by cross-validating process and module sets. Based on this detection capability, a dual-stage response strategy is employed, as shown in Fig.~\ref{fig5}.

For pointer-hijacking attacks, including tampering with the \texttt{sys\_call\_table}, \texttt{file\_operations}, or other kernel function pointers, atomic countermeasures based on physical-page attribute locking are employed. By setting page-level access permissions at the hardware level, LOOM enforces physical-layer access control, rendering critical code and data read-only or non-executable and preventing modification even by privileged kernel code. For attacks with flexible memory footprints, such as dynamically loadable kernel modules or hidden processes, an adaptive agent is deployed for active mitigation, enabling targeted and context-aware defensive actions.

\textbf{Atomic Operations.}
Hardware-enforced access control is applied directly to target physical pages via the Secure World MMU. Specifically, the Secure Monitor configures \texttt{TTBR0\_EL3} and \texttt{TCR\_EL3} in the EL3 context, while \texttt{MAIR\_EL3} defines the memory attribute encodings.  On the secure page tables, the target pages are marked as read-only through the AP field, with the PXN/UXN bits set to disable execution. The \texttt{SCR\_EL3}.NS bit governs the security state of transactions, ensuring that write operations to these pages are permitted only from the Secure World.  TLB invalidation is performed after attribute updates to guarantee immediate effect, and all modifications are completed within a critical section to preserve atomicity.  Page attributes established by the Secure World cannot be overridden by the Normal World.

\textbf{Adaptive Agents.}
During system initialization, the Secure Monitor (EL3) configures the GICv3 for secure interrupt handling.  Software Generated Interrupts (SGIs) are assigned to Group 0 (Secure Group) and given the highest priority.  The register \texttt{ICC\_IGRPEN0\_EL1} enables these interrupts, while \texttt{ICC\_PMR\_EL1} establishes a priority mask for preemption, and \texttt{GICD\_IGROUPRn} designates certain interrupts as Secure.
When a rootkit is detected, LOOM triggers a Secure SGI by writing to \texttt{ICC\_SGI0R\_EL1}, which directs control from EL1 to the EL3 FIQ handler.  This transition is hardware-enforced, safeguarding it from kernel compromise.  The Secure Monitor saves the EL1 context, updates \texttt{SCR\_EL3} to ensure Secure-state execution, and masks non-secure interrupts, effectively freezing all REE cores.  Finally, it clears pending interrupts, disconnects Redistributors, and conducts a controlled non-secure warm reset by adjusting \texttt{ELR\_EL3} and executing ERET, thereby restarting the Normal World kernel.

\subsection{Performance Optimization}

In the lightweight semantic reconstruction framework, interpreting kernel semantics requires resolving physical addresses. Each semantic query performs virtual-to-physical translation through a multi-level page table walk, which incurs significant overhead. In practice, monitoring frequently accesses the same kernel objects, such as processes, resulting in many redundant translations.

To reduce this overhead, we integrate an address translation cache into the monitoring mechanism. Kernel memory accesses exhibit strong temporal locality, as the virtual addresses of identified kernel objects are repeatedly used across monitoring iterations. By caching resolved virtual-to-physical mappings, repeated page table walks can be avoided.
The cache is implemented as a compact array-based structure with hash indexing for fast lookups. It contains 256 entries ($2^8$), balancing cache hit rate and memory overhead in the Secure World. Hash indexing is computed using a bitwise AND operation on the virtual address. Bitwise operations typically complete within a single CPU cycle, whereas arithmetic operations such as modulo or division may require tens of cycles, making this approach at least an order of magnitude more efficient. Each entry stores a virtual address tag and its corresponding physical address. On a hit, the physical address is returned directly; on a miss, a full translation is performed, and the mapping is inserted into the cache. A direct-mapped replacement policy avoids additional metadata and control overhead. Due to the high locality of kernel address accesses, this lightweight design remains effective and significantly reduces semantic reconstruction overhead.
Therefore, the cache design prioritizes minimal lookup overhead over theoretical optimality, leveraging workload characteristics to achieve a practical balance suitable for real-time kernel monitoring.

\section{Evaluation}
In this section, we evaluate the effectiveness of LOOM through experiments. Specifically, we focus on the following research questions:
\begin{itemize}
\item [$\bullet$] RQ1:What is the impact of LOOM on Normal World performance?
\item [$\bullet$] RQ2:To what extent does LOOM minimise context-switch communication overhead while maintaining secure isolation and the ability to monitor?
\item [$\bullet$] RQ3:How effectively does LOOM reduce semantic overhead via cache while maintaining rich kernel-level visibility?
\end{itemize}

\begin{table}[ht]
\centering
\caption{Configuration of hardware and software in the experiment setting.}
\label{table0}
\begin{tabular}{>{\raggedright}p{2cm}>{\raggedright}p{3.2cm}>{\raggedright\arraybackslash}p{4cm}}
\toprule
\textbf{Category} & \textbf{Item} & \textbf{Specification} \\
\midrule
\multirow{4}{*}{Hardware} & Development board & Phytium D2000 \\
& SSD & Samsung 870EVO 500GB \\
& Memory & Gloway 16GB \\
& TEE Chip & GigaDevice GD25LQ128E \\
\midrule
\multirow{3}{*}{Software} & Desktop OS &  Ubuntu 22.04 \\
& Linux Kernel & 5.10.239 \\
& OP-TEE & 3.12 \\
\bottomrule
\end{tabular}
\end{table}

\subsection{Experiment Setting}

LOOM was implemented on an ARMv8-A Phytium D2000 platform, running Linux in the Normal World and OP-TEE~\cite{ref21} in the Secure World. To support periodic out-of-band monitoring, the FreeRTOS~\cite{ref22} scheduler was ported into the OP-TEE runtime. Timer interrupts on the Secure World side were configured to drive time-slice rotation, enabling continuous execution of monitoring tasks independent of Normal World scheduling. The hardware and software configuration is summarized in Table~\ref{table0}.

\subsection{Active System Measurement(RQ1)}

\begin{figure}
\centering
\hspace*{-1cm}
\includegraphics[width=0.6\textwidth]{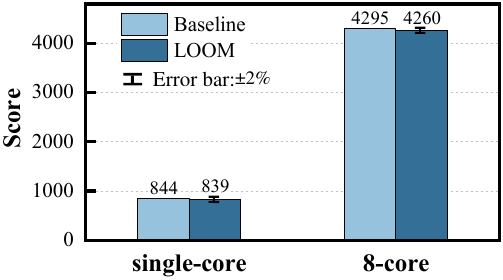}
\caption{Performance evaluation of Unixbench} \label{fig4}
\end{figure}

We use UnixBench~\cite{ref23} to evaluate whether out-of-band monitoring affects the performance of the Normal World. UnixBench includes a range of workloads, such as process creation, file-system operations, system calls, and compute-intensive tasks, to assess OS responsiveness and throughput.
To reduce measurement noise, each configuration is executed ten times, and we report the average score across all runs. As shown in Fig.~\ref{fig4}, compared with the baseline, the performance overhead is 0.6\% in single-core mode and 0.8\% in multi-core mode, both of which fall within the ±2\% error margin shown in the figure. These results indicate that LOOM introduces negligible performance overhead to Normal World execution. The monitoring logic operates in the Secure World and performs lightweight periodic kernel-state acquisition and semantic reconstruction, which minimally interferes with the execution and scheduling of the Normal World.

\subsection{Communication Overhead Analysis(RQ2)}

To answer RQ2, we evaluate the communication overhead of REE–TEE interactions using CPU cycle counts. LOOM uses a shared-memory channel to transfer kernel-critical data, avoiding frequent world switches between the REE and TEE.
Table~\ref{table1} summarizes the results. Retrieving kernel data through the shared-memory channel requires approximately $3.5 \times 10^{4}$ CPU cycles on the REE side. In contrast, performing a round-trip SMC-based context switch between the REE and TEE incurs about $7.6 \times 10^{8}$ CPU cycles, which is over four orders of magnitude higher. These results indicate that world-switch operations dominate the communication overhead in REE–TEE interactions.

\begin{table}[htbp]
  \centering
  \caption{Communication overhead of REE–TEE interaction (CPU Cycles).}
  \label{table1}
  \begin{tabular}{
    >{\raggedright\arraybackslash}p{4.8cm}
    >{\centering\arraybackslash}p{2.5cm}
    >{\centering\arraybackslash}p{4cm}
  }
    \toprule
    Operation & CPU Cycles & Reduction vs. SMC \\
    \midrule
    LOOM(with shared memory)  & $3.5 \times 10^4$ & $>10^4\times$  \\
    SMC-based context switch  & $7.6 \times 10^8$ & -- \\
    \bottomrule
  \end{tabular}
\end{table}

\subsection{Effectiveness of Semantic Overhead Reduction(RQ3)}
To evaluate whether LOOM reduces semantic reconstruction overhead while preserving kernel-level visibility, we measure the CPU cycles required for semantic parsing with and without cache-assisted translation.

The cache stores previously resolved address translations, avoiding repeated lookups during kernel-object reconstruction. We further conduct an ablation study to quantify the effectiveness of the cache mechanism. Our measurements show that semantic reconstruction without caching requires approximately $1.47 \times 10^{8}$ CPU cycles per resolution, whereas a cache miss incurs about $1.7 \times 10^{8}$ CPU cycles due to cache lookup and full page-table traversal overhead. In contrast, a cache hit requires only $2.7 \times 10^{7}$ cycles, which is approximately $6.3 \times$ faster than a cache miss. This large gap indicates that semantic overhead is primarily dominated by address translation.

The average access time with caching follows the standard cache model:

\begin{equation}
T_{\text{avg}} = H \cdot T_{\text{hit}} + (1 - H) \cdot T_{\text{miss}},
\end{equation}

where $H$ is the cache hit rate, and $T_{\text{hit}}$ and $T_{\text{miss}}$ denote hit and miss latency, respectively. Substituting the measured values gives:

\begin{equation}
T_{\text{avg}} = 1.7 \times 10^{8} - H \cdot 1.43 \times 10^{8}.
\end{equation}

Without caching, each access requires approximately $1.4 \times 10^{8}$ cycles. Caching becomes beneficial when the hit rate exceeds 0.21. Due to the spatial and temporal locality of kernel memory accesses, this condition is typically satisfied in practice.
For example, with an 80\% hit rate, the average access latency is reduced by roughly 60\%. These results demonstrate that the cache significantly reduces semantic overhead while maintaining comprehensive system visibility.
The cache is implemented as a hash table of size $2^{n}$, allowing index computation via a single AND operation. This design avoids modulo operations typically used in hash tables, which require 10–40 cycles, further reducing lookup overhead.

\subsection{Capability-oriented Comparison with SOTA works}
Due to fundamentally different execution models (e.g., native in-kernel, hypervisor-based, and TrustZone-based designs), direct numerical comparisons are often not directly meaningful. Instead, we provide a capability-oriented comparison that aligns key dimensions such as isolation, semantic visibility, overhead, and trusted computing base (TCB)/attack surface, which more accurately reflects the underlying design trade-offs. The result is shown in Table~\ref{table2}.

\begin{table}[htbp]
\centering
\caption{Capability-oriented comparison with SOTA works.}
\label{table2}
\renewcommand{\arraystretch}{1.3}
\setlength{\tabcolsep}{4pt}

\begin{tabular}{
>{\centering\arraybackslash}m{1.5cm}|
>{\centering\arraybackslash}m{1.4cm}|
>{\centering\arraybackslash}m{1.8cm}|
>{\centering\arraybackslash}m{1.5cm}|
>{\centering\arraybackslash}m{3.2cm}|
>{\centering\arraybackslash}m{1.1cm}
}

\hline

\textbf{Approach} 
& \textbf{Isolation} 
& \textbf{Semantic \newline Visibility} 
& \textbf{Overhead} 
& \textbf{TCB \newline (Attack Surface)} 
& \textbf{Active \newline Defense} \\
\hline
In-kernel 
& Low 
& High 
& Moderate 
& Kernel OS \newline (easily attackable) 
& No \\
\hline

Hypernel \newline (hardware-assisted)
& High 
& Low 
& Low 
& Hardware \newline (side-channel risk) 
& No \\

\hline

NFM (VMI) 
& Medium 
& High 
& Negligible 
& Hypervisor \newline (complex attack surface) 
& No \\

\hline

KIMS (TZ) 
& High 
& Medium 
& Low 
& TZ secure world \newline (minimal exposure) 
& No \\

\hline

\textbf{LOOM (TZ)} 
& \textbf{High} 
& \textbf{Semantic-aware} 
& \textbf{Negligible} 
& \textbf{TZ secure world} \newline \textbf{(minimal exposure)} 
& \textbf{Yes} \\
\hline
\end{tabular}
\end{table}

\section{Security Capability Analysis}
We analyze the defensive capability of LOOM and show its effectiveness using representative CVE (Common Vulnerabilities and Exposures) cases~\cite{ref24}. 
LOOM is designed to ensure the semantic correctness of security-critical kernel states, rather than targeting a single attack instance or exploit chain.  Therefore, the selected CVEs represent broader exploit classes, such as privilege escalation, process hiding, and syscall hijacking, covering common kernel-level attack behaviors while remaining representative of real-world attack scenarios.

By monitoring critical kernel objects and reconstructing their semantics, LOOM provides fine-grained supervision of kernel behavior. As summarized in Table~\ref{table3}, we focus on three security-critical kernel objects that are frequently targeted in attacks: the process descriptor (\texttt{init\_task}), process credentials (\texttt{init\_cred}), and the system call table (\texttt{sys\_call\_table}). The following examples illustrate how LOOM links semantic monitoring of critical kernel objects to real-world vulnerability scenarios. 

\begin{table}[htbp]
\centering
\caption{The security capability analysis based on the CVE cases.}\label{table3}
\renewcommand{\arraystretch}{1.3}
\setlength{\tabcolsep}{4pt}
\begin{tabular}{
>{\centering\arraybackslash}m{1.75cm}|
>{\centering\arraybackslash}m{2cm}|
>{\centering\arraybackslash}m{1.5cm}|
>{\centering\arraybackslash}m{1.5cm}|
>{\centering\arraybackslash}m{3.3cm}
}
\hline
\textbf{Description} & \textbf{Objects} & \textbf{Physical \newline address} & \textbf{Semantic \newline Info} & \textbf{CVE Cases} \\
\hline

Process \newline Information & init\_task & 0xf3f12a00 & pid, \newline state, \newline comm, \newline process list &
\textbf{CVE-2021-22555} \newline(Netfilter) \newline
\textbf{CVE-2005-0736} \newline(Process Hiding) \newline
\textbf{CVE-2021-22668} \newline(Privilege Escalation) \\
\hline

Process \newline Credentials & init\_cred & 0xf3f1dd88 & cred, \newline count, \newline user IDs &
\textbf{CVE-2021-33909} \newline(Qualys Serial Overflow) \newline
\textbf{CVE-2022-2588} \newline(Use-After-Free) \newline
\textbf{CVE-2021-4154} \newline(Use-After-Free) \\
\hline

System Call  & sys\_call\_table & 0xf3700840 & Function \newline pointers &
\textbf{CVE-2010-3301} \newline(Privilege Escalation) \newline
\textbf{CVE-2016-5195} \newline(Dirty COW) \newline
\textbf{CVE-2017-5123} \newline(waitid Vulnerability) \\
\hline

\end{tabular}
\end{table}

For process information, LOOM reconstructs semantics from the physical address of \texttt{init\_task}, including the process ID (PID), task state, executable name (comm), and task list pointers. By verifying task-list integrity and detecting abnormal changes in the comm field, LOOM can identify attacks such as process hiding (CVE-2005-0736~\cite{ref26}), Netfilter exploitation (CVE-2021-22555~\cite{ref25}), and process injection vulnerabilities (CVE-2021-22668).

In terms of process credentials, LOOM derives its semantics from \texttt{init\_cred}, which includes credential pointers, reference counts, and user IDs (uid/euid). This information is crucial for detecting privilege escalation. For example, CVE-2021-33909~\cite{ref27} can be identified by monitoring invalid references to \texttt{init\_cred}. Additionally, use-after-free vulnerabilities, such as CVE-2022-2588~\cite{ref28} and CVE-2021-4154~\cite{ref29}, are mitigated through careful tracking of lifecycle and reference counts. Sudden changes in uid also indicate potential escalation attempts.

About the system call table, LOOM reconstructs and monitors syscall function pointers from the physical address of \texttt{sys\_call\_table}. By verifying pointer integrity and memory protection attributes, LOOM can detect kernel tampering and rootkit behavior. For example, CVE-2010-3301~\cite{ref30} can be detected through syscall pointer integrity checks. For CVE-2016-5195~\cite{ref31}, LOOM monitors write-protection of the syscall table page and detects illegal writes via atomic hazard prevention. In the case of CVE-2017-5123~\cite{ref32}, abnormal modifications of syscall pointers reveal malicious hooks.

\section{Conclusion}
In this paper, we present LOOM, an ARM TrustZone-based lightweight out-of-band OS monitor for kernel security. By locating monitoring logic in the Secure World, LOOM eliminates the shared-fate dependency between protection mechanisms and a potentially compromised kernel, thereby providing stronger isolation. We further introduce a lightweight semantic reconstruction approach that selectively captures the state of critical kernel objects, enabling accurate security analysis while minimizing semantic overhead. Additionally, a dual-stage hazard prevention strategy provides both proactive protection for sensitive kernel memory regions and rapid response to malicious kernel behaviors. The prototype implementation and experimental evaluation show that LOOM brings negligible performance overhead. Overall, LOOM provides an efficient and practical foundation for hardware–software co-designed secure kernel monitoring.

\begin{credits}
\subsubsection{\ackname} This work was supported in part by the National Natural Science Foundation of China under Grant No.62472086, and in part by the Fundamental and Interdisciplinary Disciplines Breakthrough Plan of the Ministry of Education of China under Grant No. JYB2025XDXM118.

\end{credits}

\end{sloppypar}
\end{document}